\documentclass[11pt]{article}
\usepackage{graphicx}
\usepackage{amsfonts,ulem}
\usepackage{amsmath}
\usepackage{amssymb}
\usepackage{color}
\usepackage{braket}
\usepackage[utf8]{inputenc}
\usepackage{hyperref} 
\usepackage{subcaption}
\usepackage{float}

\pdfoutput=1

\def \be {\begin{equation}}
\def \ee {\end{equation}}
\def \ba {\begin{aligned}}
\def \ea {\end{aligned}}
\def \bea {\begin{eqnarray}}
\def \eea {\end{eqnarray}}

\begin{document}
%if0
\begin{titlepage}
\begin{flushright}
%NORDITA 2020-??? \\
October, 2024
\end{flushright}
\vspace{0.5cm}
\begin{center}
{\Large \bf Truncating Dyson-Schwinger Equations Based on Lefschetz Thimble Decomposition and Borel Resummation
}
\lineskip .75em
\vskip 2.5cm
{Feiyu Peng$^{a,}$\footnote{pengfy@gs.zzu.edu.cn} and Hongfei Shu$^{a,b,c,}$\footnote{shuphy124@gmail.com, shu@zzu.edu.cn}
}
\vskip 2.5em
 {\normalsize\it 
$^{a}$Institute for Astrophysics, School of Physics,
Zhengzhou University, Zhengzhou, Henan 450001, China\\
$^{b}$Beijing Institute of Mathematical Sciences and Applications, Beijing, 101408, China\\
$^{c}$Yau Mathematical Sciences Center, Tsinghua University, Beijing, 100084, China
}
\vskip 3.0em
\end{center}
\begin{abstract}
%Truncating the Dyson-Schwinger (DS) equations in quantum field theory presents a practical challenge. 
%In this paper, 
We study the zero-dimensional prototype of the path integrals in quantum mechanics and quantum field theory, with the action $S(\phi)=\frac{\sigma }{2}\phi ^{2}  +\frac{\lambda}{4} \phi ^{4}$. Using the Lefschetz thimble decomposition and the saddle point expansion, we derive multiple asymptotic formal series of the correlation function associated with the perturbative and non-perturbative saddle points. Furthermore, we reconstruct the exact correlation function employing the Borel resummation. We then consider how to truncate the Dyson-Schwinger (DS) equations beginning with the perturbation expansion of the correlation functions, analogous to the one obtained from the Feynmann diagram in higher dimensions. For the case $\sigma<0$, we find that although the asymptotic
series around the perturbative saddle point is Borel summable, it does not capture the full information. Consequently, contributions from non-perturbative saddle points must be included to ensure a complete truncation procedure.

  \end{abstract}
\end{titlepage}

\section{Introduction}
The development of non-perturbative methods is one of the most critical tasks in Quantum Mechanics (QM) and Quantum Field Theory (QFT). Among the recent advancements, the resurgence theory has drawn much attention. A general feature of many interesting QFTs is that the perturbative series for quantities such as correlation function and scattering amplitude are always divergent \cite{Dyson:1952tj,GZ-1990}. Resurgence theory provides a powerful method to convert these divergent series to meaningful objects and reconstruct the original quantities via the Borel resummation. See \cite{Marino:2012zq,Dorigoni:2014hea,Aniceto:2018bis} for nice reviews and references therein. Moreover, the resurgence theory has uncovered a mysterious connection between the non-perturbative effects and the perturbation theory, which provides a promising framework for developing the non-perturbative methods. 

Another classic non-perturbative method is the Dyson-Schwinger (DS) equations \cite{Dyson:1949ha,Schwinger:1951ex,Schwinger:1951hq}, which provide a formalism to compute Green's functions (or correlation functions) in QFT and are particularly useful in the study of quantum chromodynamics \cite{Roberts:1994dr,DSE-2000}. Since the DS equations provide an infinite tower of equations, one has to truncate it to a finite system 
in practice. However, a challenge arises due to the underdetermination of the Dyson-Schwinger equation \cite{Bender:1988bp}, because there are more correlation functions than the equations. To address this issue, a common approach is to set the highest correlation functions to zero, which is denoted as the traditional truncation in this paper. See \cite{Okopinska:1990pt} for example. However, this method usually provides values of correlation functions that differ from the exact ones, as correlation functions do not approach zero as $n$ increases. Alternatively, the other truncation based on the asymptotic behavior of the correlation function has been proposed in \cite{Bender:2022eze,Bender:2023ttu}, which has shown a dramatic improvement in approximating the DS equations. 

In this paper, we study the truncation of the DS equations based on the exact results obtained from the resurgence theory.
In particular, we will consider the zero-dimensional
prototype of the path integrals in QM and QFT, $S(\phi)=\frac{\sigma }{2}\phi ^{2}  +\frac{\lambda}{4} \phi ^{4}$. When $\sigma>0$ ($\sigma<0$), this corresponds to the anharmonic oscillator (double well potential). This model serves as one of the simplest but non-trivial toy models in QM and QFT\footnote{Some recent progress on the cubic and quartic toy model can be found in \cite{Ivanov:2021yag,Konosu:2024zrq,Edery:2024chj,Grimm:2024hdx}.}. Since the perturbation series is the main object in most calculations of QFT, we will also start with the perturbation expansion of the correlation functions of the zero-dimensional QFT to gain insights into more general QFT\footnote{This perturbation expansion is an analogue to the one obtained from the Feynmann diagram in general QFT.}. Due to the presence of the multiple saddle points in $S(\phi)$, one should use Lefschetz thimble decomposition \cite{Witten:2010cx} to see which saddle points contribute to the integrals. We find only one (real) saddle point contributes as $\sigma>0$, while three saddle points for $\sigma<0$. We then consider the saddle point expansion of the correlation functions, which is clearly divergent. By using the Borel resummation, we reconstruct the exact results of the (connected) correlation functions. Based on these exact results, we propose a truncation method of the DS equations by using the asymptotic behavior of the (connected) correlation functions, which shows a greater improvement than the traditional one.

This paper is organized as follows: in section \ref{sec:zero-QFT-DS}, we introduce the zero-dimensional QFT, and derive the DS equations for the (connected) correlation functions. In section \ref{sec:resurgence}, we first consider the Lefschetz thimble decomposition of the correlation function, then show how to get the exact result by using the Borel resummation. In section \ref{sec:truncation}, we will propose a truncation of the DS equations beginning from the asymptotic behavior, and compare it with the traditional one. In section \ref{sec:conclusion}, we provide a summary of our results and make several remarks on the truncation of the DS equation, which may shed light on the general QFTs. Possible future directions will also be shown.

\section{A zero-dimensional QFT model and DS equations}\label{sec:zero-QFT-DS}
In this section, we introduce the (connected) correlation function of QFT and derive their DS equations. We will focus on the following zero-dimensional prototype of the path integrals in QM and QFT
  \begin{equation}
S\left [ \phi  \right ] =\frac{\sigma }{2}\phi ^{2}  +\frac{\lambda}{4} \phi ^{4}-J\phi,
\end{equation}
where we have introduced a source term $J\phi$ in the action. When $\sigma>0$ ($\sigma<0$), it corresponds to the anharmonic oscillator (double well potential). The corresponding Euclidean partition function reads
\begin{equation}
Z[J]=\int_{-\infty}^{\infty}d\phi \ e^{-S[\phi]}=\int_{-\infty}^{\infty}d\phi \ e^{-\frac{\sigma }{2}\phi ^{2}  -\frac{\lambda}{4} \phi ^{4}+J\phi }.
\end{equation}
It is the generating function of the (nonconnected) correlation function: 
\begin{equation}
Z[J]/Z[0]=\sum_{n=0}^{\infty}\frac{\gamma_n}{n!}J^n,
    \label{Z exp}
\end{equation}
where $\gamma_n$ is \footnote{Since we are considering the zero-dimensional QFT, the functional derivative $\delta$ has been replaced by the derivative $\partial$.}
\begin{equation}
\gamma _n\!=\!\frac{1}{Z[0]}\frac{\partial^n }{\partial J^n}Z[J] \Big |_{J\equiv0}.
\label{gamma definition}
\end{equation}
Here we have used the normalized partition function.
%, where $Z[0]$ represents the partition function with $J=0$. 
Due to the parity invariance of the theory, the correlation function with odd fields always vanishes, $\gamma_{2n+1}=0$. The non-trivial correlation function is 
\begin{equation}
\gamma _{2n}=\frac{\int_{-\infty}^{\infty}d\phi \ \phi^{2n}e^{-\frac{\sigma }{2}\phi ^{2}  -\frac{\lambda}{4} \phi ^{4} }}{\int_{-\infty}^{\infty}d\phi \ e^{-\frac{\sigma }{2}\phi ^{2}  -\frac{\lambda}{4} \phi ^{4} }},
\label{gamma 2n}
\end{equation}
where $n=1,2,\cdots$.
 The quantum version of the equation of motion of this model is given by
\begin{equation}
\sigma\frac{\partial Z[J]}{\partial J}+\lambda\frac{\partial^3 Z[J]}{\partial J^3}-J Z[J]=0.
\label{Z ode}
\end{equation}
Substituting \eqref{Z exp} into this differential equation, one obtains the DS equations of the correlation functions:
\begin{equation}\label{eq:DS-non-dis}
\begin{aligned}
\gamma_4\!&=\!-\frac{\sigma}{\lambda}\gamma_2+\frac{1}{\lambda},\\
\gamma_6\!&=\!-\frac{\sigma}{\lambda}\gamma_4+\frac{3}{\lambda}\gamma_2,\\
\gamma_8\!&=\!-\frac{\sigma}{\lambda}\gamma_6+\frac{5}{\lambda}\gamma_4,\\
\!&\cdots\!\,.
\end{aligned}
\end{equation}
The connected correlation function is 
\begin{equation}
G _n=\frac{\partial^n }{\partial J^n}\log{Z}[J] \Big |_{J\equiv0}.
\label{G difin}
\end{equation}

Let $W[J]\equiv\log{Z[J]}$, quantum equation of motion (\ref{Z ode}) can be written as a differential equation of $W[J]$:

\begin{equation}
\lambda \frac{\partial^3}{\partial J^3}W[J]+3\lambda\frac{\partial}{\partial J}W[J]\frac{\partial^2}{\partial J^2}W[J]+\lambda\left(\frac{\partial}{\partial J}W[J]\right)^3+\sigma \frac{\partial}{\partial J} W[J]=J.
\label{odeW}
\end{equation}
Substituting $W[J]=\sum_{n=0}^{\infty}\frac{G_{2n}}{(2n)!}J^{2n}$ into the above differential equation, one finds the DS equations of the connected correlation function:
\begin{equation}\label{eq:DS-conn}
\begin{aligned}
G_4\!&=\!-\frac{ \sigma }{\lambda }G_2-3 G_2^2+\frac{1}{\lambda },\\
G_6\!&=\!-\frac{\sigma }{\lambda }G_4 -6 G_2^3-12 G_4 G_2,\\
G_8\!&=\!-\frac{\sigma }{\lambda }G_6 -60 G_4 G_2^2-18 G_6 G_2-30 G_4^2,\\
\!&\cdots\!\,.
\end{aligned}
\end{equation}

As shown in \eqref{eq:DS-non-dis} and \eqref{eq:DS-conn}, the DS equations include an infinite tower of equations. Usually, one will truncate the infinite system of DS equations into finite number equations.
However, as noted in \cite{Bender:1988bp}, the DS equations are undetermined, because the (connected) correlation functions always outnumber the equations. To resolve this issue, one usually sets the highest correlation functions to zero, which is called the traditional truncation. However, this method usually provides different values of correlation functions. The other truncation based on the asymptotic behavior of the correlation function has been proposed in \cite{Bender:2022eze} for the case $\sigma=0$.

In the following of this paper, we will start with the perturbative expansion of the correlation function, and show how to compute it exactly in the context of resurgence theory. Based on this result, we will propose a truncation method of the DS equations.

\section{Saddle point method and resurgence theory}\label{sec:resurgence}
In this section, we derive the perturbative expansion of the correlation function by using Lefschetz thimble decomposition and saddle point expansion, and then show how to compute the correlation function exactly in the context of resurgence theory.

\subsection{Lefschetz thimble and saddle point expansion
}
Let us first consider the following integral 
\begin{equation}
    F=\int_{\mathcal{C}}dz\ g(z)e^{-\frac{1}{\lambda}f(z)},
\end{equation}
where $z$ and $\lambda$ are complex in general. To compute this integral exactly, we consider its Lefschetz thimble decomposition \footnote{More details on the Lefschetz thimbles can be found in \cite{Witten:2010cx}. See also \cite{Serone:2017nmd}. }. 
We assume all the saddle points $z_i$, $\delta f(z_i)/\delta z=0$, are isolated and non-degenerate, $\partial^2_zf\neq 0$. The Lefschetz thimble is defined by the following flow equation emerging from the saddle point $z_i$
\begin{equation}
\frac{d z(t)}{d t}=a\overline{\left(\frac{\partial f(z)/\lambda}{\partial z}\right)},
\label{flow eq}
\end{equation}
where $t$ parameterizes the thimble and the overline means taking the complex conjugate. $a=1$ corresponds to the Lefschetz thimble $\mathcal{J}_i$ (downward flow), and $a=-1$ corresponds to the anti-thimble $\mathcal{K}_i$ (upward flow). The imaginary part of $-f/\lambda$ is constant on the thimble ${\cal J}_i$ and ${\cal K}_i$, while the real part of $-f/\lambda$ will always decrease (increase) along the flow ${\cal J}_i$ (${\cal K}_i$), which guarantees the convergence of the integral along the thimble ${\cal J}_i$. 

Suppose there are no singularities on the complex plane, the contour ${\cal C}$ can be decomposed into thimbles ${\cal C}=\sum_in_i{\cal J}_i$, such that
\begin{equation}
    F=\sum_i n_iF_i\equiv\sum_i n_i\int_{\mathcal{J}_i}dz\ g(z)e^{-\frac{1}{\lambda}f(z)},
\end{equation}
where coefficients $n_i$ is given by the intersection number $n_i=\langle {\cal C}, {\cal K}_i \rangle$.

Unfortunately, the exact calculation of the integral along the thimble is not known. Usually, we approximate each integral by using the saddle point expansion
\begin{equation}
    F_i=%e^{-\frac{1}{\lambda}f(x_i)}
    \sum_{k=0}^\infty F_i^{(k)} \lambda^k.
\end{equation}
This expansion is an asymptotic formal series and should be reconstructed by using Borel resummation. We will use a generalization of the Borel transform, called as Borel-Le Roy transform, defined by\footnote{Note that the location of the singularity of Borel-Le Roy transform is independent of parameter $b$.}
\begin{equation}
    \mathcal{B}_b[F_i](s)=\sum_{k=0}^{\infty}\frac{F_i^{(k)}}{\Gamma(k+1+b)}s^k,
\label{Borel trans}
\end{equation}
where $s$ is the coordinate of the Borel plane, $b$ is a real parameter that can be chosen as needed.
The Borel resummation is performed by the Laplace transformation
 \begin{equation}
\mathcal{S}[F_i](\lambda)=\int_0^{\infty}dt\ t^be^{-t}\mathcal{B}_b[F_i](\lambda t).
\label{Borel resum}
\end{equation}

If no singularity of the Borel transform lies on the integration path of Borel resummation, the asymptotic series $F_i$ is Borel summable, otherwise, $F_i$ is Borel non-summable. For non-summable cases, we will provide approaches in specific examples to address the issue.

\subsection{Lefschetz thimble decomposition for zero-dimensional QFT}
We now apply the Lefschetz thimble decomposition to the unnormalized correlation function 
\begin{equation}
{\cal I}_{2n}\equiv\int_{-\infty}^{\infty}d\phi\ \phi^{2n}e^{-S(\phi)}.
\label{A unnorm}
\end{equation}
It is useful to rescale the field by  $x=\sqrt{\lambda}\phi$, such that the unnormalized correlation function becomes \footnote{More precisely, we have complexified the $x$ and $\lambda$. }
\begin{equation}
  {\cal I}_{2n}\!  \!=\!\frac{1}{\sqrt{\lambda}}\lambda^{-n}\int_{{\cal C}_R}dx\ x^{2n}e^{-\frac{1}{\lambda}\hat{S} (x)},\quad \hat{S} (x)=\frac{\sigma}{2}x^2+\frac{1}{4}x^4,
\end{equation}
where ${\cal C}_R$ denotes the integral on real axis.
We have three real saddle points for $\sigma<0$, and only one real saddle point for $\sigma>0$
\begin{equation}
    \begin{cases}
x_{0}=0,x_{\pm}=\pm i\sqrt{\sigma} & \sigma>0\\
x_{0}=0,x_{\pm}=\pm\sqrt{-\sigma} & \sigma<0
\end{cases}.
\end{equation}

\begin{figure}[t]
\begin{center}
\begin{subfigure}{0.28\textwidth}
\includegraphics[width=\linewidth]{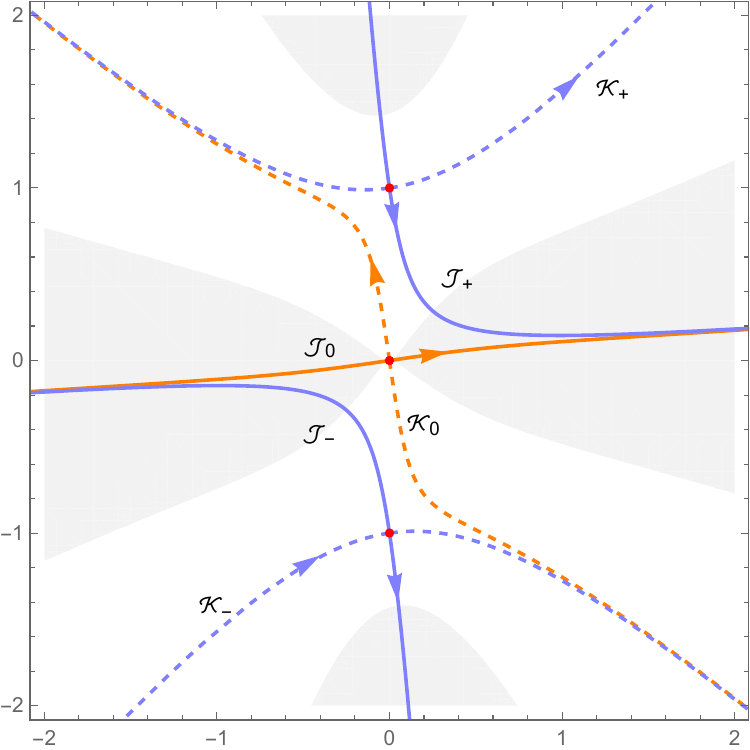}
    \caption{}
    \end{subfigure}
    \begin{subfigure}{0.28\textwidth}
 \includegraphics[width=\linewidth]{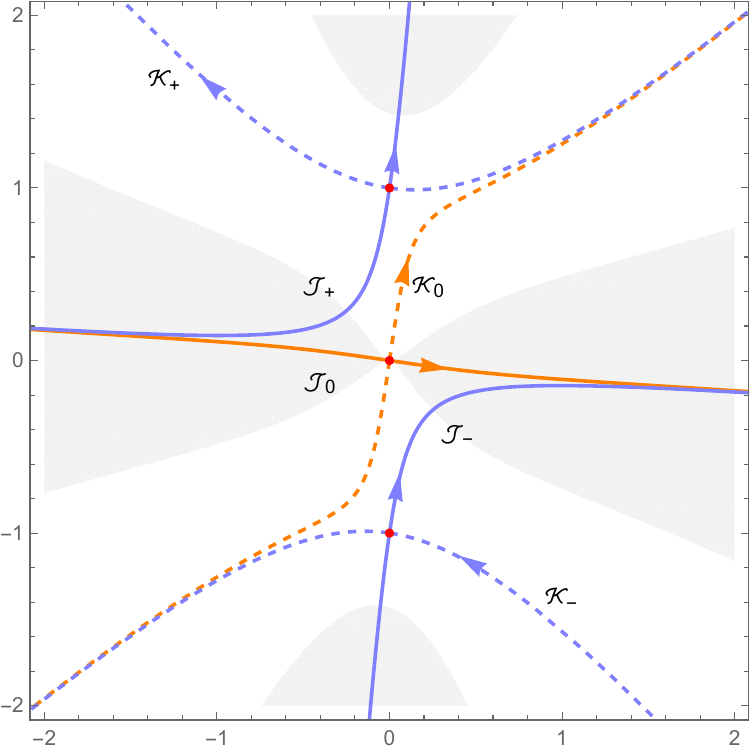}
    \caption{}     
    \end{subfigure} \\ 
    \begin{subfigure}{0.28\textwidth}
\includegraphics[width=\linewidth]{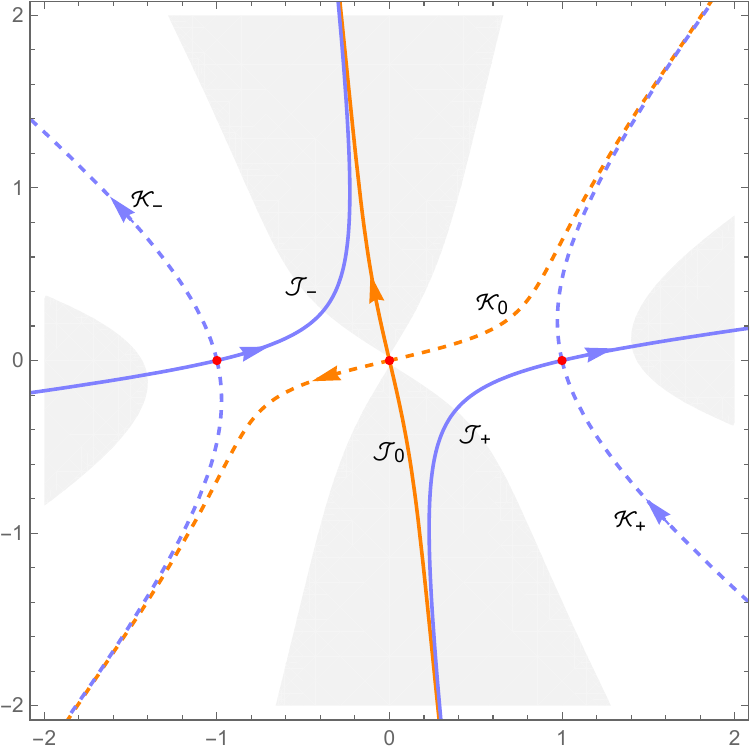}
    \caption{}
    \end{subfigure}
    \begin{subfigure}{0.28\textwidth}
 \includegraphics[width=\linewidth]{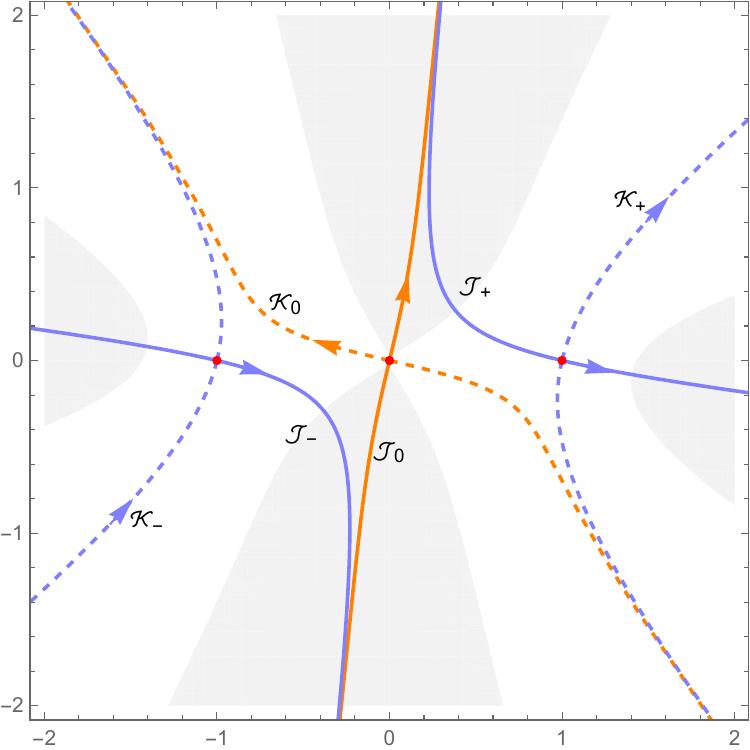}
    \caption{}     
    \end{subfigure} 
    \end{center}
    \caption{The upper panels (lower panels) of the figure correspond to $\sigma>0$ ($\sigma<0$), while the left panels (right panels) correspond to ${\rm Im}\lambda>0$ (${\rm Im}\lambda<0$). The solid curves (dashed curves) represent the Lefschetz thimbles (anti-thimbles). The red dots represent the saddle points. Orange curves (blue curves) correspond to the thimbles of $x_0$ ($x_\pm$). The shaded regions represent the regions where the integral converges.}
    \label{fig:thimble}
\end{figure}
Here $x_0$ and $x_\pm$ are called the perturbative saddle point and non-perturbative saddle point respectively.
The thimbles for the action $S(x)$ is shown in Fig.\ref{fig:thimble}.\footnote{To solve the flow equation numerically, we start with a neighbourhood of the saddle point, determined by the Hessian matrix, and then solve it using ``DSolve'' in Mathematica. See \cite{Alexandru:2020wrj} for reviews and references therein.} One thus can find the decomposition of the contour ${\cal C}_R$. When $\sigma>0$, ${\cal C}_R$ is nothing but the thimble ${\cal J}_{0}$:
\begin{equation}\label{contour_sigma>0}
    {\cal C}_{R}={\cal J}_{0},\quad\sigma>0.
\end{equation}

When $\sigma<0$, ${\cal K}_\pm$ will also intersect with the real axis, one finds
\begin{equation}
    {\cal C}_{R}=\begin{cases}
{\cal J}_{-}-{\cal J}_{0}+{\cal J}_{+} & {\rm Im}(\lambda)>0\\
{\cal J}_{-}+{\cal J}_{0}+{\cal J}_{+} & {\rm Im}(\lambda)<0
\end{cases},\quad \sigma<0,
\label{contour}
\end{equation}
where the orientation of the thimble is shown in Fig.\ref{fig:thimble}. The integral ${\cal I}_{2n}$ thus decompose into
\begin{equation}
    {\cal I}_{2n}=\sum_{i}n_{i}{\cal I}_{2n.i}=\sum_{i}n_{i}\int_{{\cal J}_{i}}x^{2n}e^{-\frac{1}{\lambda}\hat{S} (x)}.
\end{equation}

Each integral ${\cal I}_{2n,i}$ can be approximated by the saddle point expansion
\begin{equation}\label{eq:I-saddle-Exp}
    {\cal I}_{2n.i}=e^{-\frac{1}{\lambda}\hat{S} (x_{i})}\lambda^{-n}\int_{-e^{-i\frac{\theta}{2}}\infty}^{e^{-i\frac{\theta}{2}}\infty}dq\ (\sqrt{\lambda}q+x_{i})^{2n}e^{-\frac{1}{\lambda}(\hat{S} (x_{i}+\sqrt{\lambda}q)-\hat{S} (x_{i}))},
\end{equation}
where $\hat{S}(x_i)$ is the on-shell action at the saddle point $x_i$
and $\theta={\rm Arg}(\lambda)$. The integration contour is chosen such that $\sqrt{\lambda} q$ is real. The integrals can be evaluated by using the Gaussian integral, which provides the asymptotic formal expansion with respect to $\lambda$.

\subsection{Correlation function}
In this subsection, we consider the integral \eqref{eq:I-saddle-Exp} for each saddle point, which leads to the asymptotic series with respect to $\lambda$. We then perform the Borel resummation, and compute the correlation function ${\cal I}_{2n}$ exactly.

\subsubsection{$\sigma>0$}
We first consider the case $\sigma>0$, where only the saddle point $x_0=0$ contributes, see \eqref{contour_sigma>0}. Expand \eqref{eq:I-saddle-Exp} at $\lambda=0$ with $x_i=x_0$ and integrate over $q$, we obtain
\begin{equation}
\tilde{\cal I}_{2n}=\sum_{k=0}^{\infty}\tilde{\cal I}_{2n,0}^{(k)}\lambda^k
=2^{n+\frac{1}{2}}\sum_{k=0}^{\infty}(-1)^k\sigma^{-2k-\frac{1}{2}-n}\frac{\Gamma\left(2k+\frac{1}{2}+n\right)}{k!}\lambda^k,
\label{A0k}
\end{equation}
where we have denoted the asymptotic formal series by $\tilde{\cal I}_{2n}$ to distinguish it with the exact form.
Perform the Borel transform with $b = -1$, one finds
\begin{equation}
\mathcal{B}_{-1}[\tilde{\cal I}_{2n}-\tilde{\cal I}_{2n}^{(0)}](s)=-2^{n+\frac{1}{2}} s \sigma^{-n-\frac{5}{2}}\Gamma\left(n+\frac{5}{2}\right)\ _2F_1\left(\frac{5}{4}+\frac{n}{2},\frac{7}{4}+\frac{n}{2},2,-\frac{4s}{\sigma^2} \right),
\end{equation}
 where ${}_2F_1 (a, b, c|z)$ is the hypergeometric function. Note that in this calculation, we have dropped the term with $k=0$, which will be added later. The Borel resummation and the Lefschetz
thimble decomposition \eqref{contour_sigma>0} lead to
\begin{equation}
\begin{aligned}
{\cal I}_{2n}(\lambda)\!&=\!\tilde{\cal I}_{2n}^{(0)}+\int_0^{\infty}dt\ t^{-1}e^{-t}\mathcal{B}_{-1}[\tilde{\cal I}_{2n}-\tilde{\cal I}_{2n}^{(0)}](\lambda t)\\
\!&=\!\lambda ^{-\frac{n}{2}-\frac{1}{4}} \Gamma \left(n+\frac{1}{2}\right) U\left(\frac{n}{2}+\frac{1}{4},\frac{1}{2},\frac{\sigma ^2}{4
   \lambda }\right),
   \label{SAlambda sigma>0}
\end{aligned}
\end{equation}
where $U$ is the Tricomi confluent hypergeometric function. Expanding the above expression (\ref{SAlambda sigma>0}) at $\lambda=0$, one reproduces the asymptotic series (\ref{A0k}).

\subsubsection{$\sigma<0$}
In this case, we need to consider all the saddle points $x_0=0, x_\pm =\pm \sqrt{-\sigma}$. At first, we consider the saddle point $x_0$:
\begin{equation}
\tilde{\cal I}_{2n,0}\!=\!\sum_{k=0}^{\infty}\tilde{\cal I}_{2n,0}^{(k)}\lambda^k
\!=\!\frac{i2^{n+\frac{1}{2}}}{\sqrt{-\sigma}}\sum_{k=0}^{\infty}(-1)^k\sigma^{-2k-n}\frac{\Gamma\left(2k+\frac{1}{2}+n\right)}{k!}\lambda^k.
\label{A00k}
\end{equation}
The Borel transform with $b=-1$ leads to
\begin{equation}
\mathcal{B}_{-1}[\tilde{\cal I}_{2n,0}-\tilde{\cal I}_{2n,0}^{(0)}](s)\!=\!i 2^{n+\frac{1}{2}} s \sqrt{-\sigma}\sigma^{-n-3}\Gamma\left(n+\frac{5}{2}\right)\ _2F_1\left(\frac{5}{4}+\frac{n}{2},\frac{7}{4}+\frac{n}{2};2;-\frac{4s}{\sigma^2} \right).
\end{equation}
It is worth noting that the singularity $s=-\frac{\sigma^2}{4}$ is not on the integral contour of the resummation, which means this asymptotic series is still Borel summable. The resummation thus can be performed normally:
\begin{equation}
\begin{aligned}
{\cal S}[{\cal I}_{2n,0}](\lambda)\!&=\!\tilde{\cal I}_{2n,0}^{(0)}+\int_0^{\infty}dt\ t^{-1}e^{-t}\mathcal{B}_{-1}[\tilde{\cal I}_{2n,0}-\tilde{\cal I}_{2n,0}^{(0)}](\lambda t)\\
\!&=\!i (-1)^n \lambda ^{-\frac{n}{2}-\frac{1}{4}} \Gamma \left(n+\frac{1}{2}\right) U\left(\frac{n}{2}+\frac{1}{4},\frac{1}{2},\frac{\sigma ^2}{4 \lambda }\right).
\label{x0 resum sigma m}
\end{aligned}
\end{equation}
Although it is Borel summable, the Borel resummation at this saddle point does not provide the complete result of the correlation function. 

We then consider the other two saddle points $x_\pm$:
\begin{equation}
\tilde{\cal I}_{2n,\pm}\!=\!\sum_{k=0}^{\infty}\tilde{\cal I}_{2n,\pm}^{(k)}\lambda^{k-n}=e^{\frac{\sigma^2}{4\lambda}}\frac{\sqrt{\pi}(1-2n)}{2\Gamma\left(\frac{3}{2}-n\right)}\sum_{k=0}^{\infty}\frac{\Gamma\left(2k+\frac{1}{2}-n\right)}{(-\sigma)^{2k+\frac{1}{2}-n}k!}\lambda^{k-n}.
\label{Apmk}
\end{equation}
The power of $\lambda$ starts from $-n$, so we perform the Borel transform on $\lambda^n \tilde{\cal I}_{2n(\pm)}$, and then multiply back the factor $\lambda^{-n}$ at the end:

\begin{equation}
\mathcal{B}_{-1}[\lambda^n (\tilde{\cal I}_{2n,\pm}-\tilde{\cal I}_{2n,\pm}^{(0)})](s)=\frac{1}{4} \sqrt{\pi } (4n^2-8n+3) e^{\frac{\sigma ^2}{4 \lambda }}s (-\sigma )^{n-\frac{5}{2}} \,
   _2F_1\left(\frac{5}{4}-\frac{n}{2},\frac{7}{4}-\frac{n}{2};2;\frac{4 s}{\sigma ^2}\right).
\end{equation}
Here the hypergeometric function ${}_2F_1$ is singular at $s=\frac{\sigma^2}{4}$ with a branch cut for $s\in [\frac{\sigma^2}{4},+\infty]$. In this case, we can add a small imaginary part to $\lambda$ to avoid this singularity in the Borel resummation. However, the Stokes phenomenon occurs during the change from ${\rm Im}\lambda>0$ to ${\rm Im}\lambda<0$, which leads to the ambiguity of the Borel resummation
\begin{equation}
\begin{aligned}
    \mathcal{S}_\pm[\tilde{\cal I}_{2n,+}](\lambda)&=\mathcal{S}_\pm[\tilde{\cal I}_{2n,-}](\lambda)=\lambda^{-n} \mathcal{S}_\pm[\lambda^n\tilde{\cal I}_{2n,+}](\lambda)\\
    &=\pm\frac{\sqrt[4]{-1} \sqrt{\pi } (\mp i)^n \lambda ^{-\frac{n}{2}-\frac{3}{4}} \Delta_\pm}{2 \Gamma \left(\frac{1}{2}-n\right)},
\end{aligned}
\end{equation}
where ${\cal S}_\pm$ denote the Borel resummation for ${\rm Im}(\lambda)>0$ and ${\rm Im}(\lambda)<0$, $\Delta_\pm$ is defined by the Kummer confluent hypergeometric function $_1F_1$
\begin{equation}
    \Delta_{\pm}=i^{\frac{1}{2}(1\mp1)}\Big[\sqrt{\lambda}\Gamma\left(\frac{1}{4}-\frac{n}{2}\right)\,_{1}F_{1}\left(\frac{1}{4}+\frac{n}{2};\frac{1}{2};\frac{\sigma^{2}}{4\lambda}\right)\pm i\sigma\Gamma\left(\frac{3}{4}-\frac{n}{2}\right)\,_{1}F_{1}\left(\frac{3}{4}+\frac{n}{2};\frac{3}{2};\frac{\sigma^{2}}{4\lambda}\right)\Big].
\end{equation}
This ambiguity will be cancelled by taking all saddle points into account through the method of Lefschetz thimbles \eqref{contour}
\begin{equation}\label{resum iml pm}
    \begin{aligned}
        {\cal I}_{2n}(\lambda)\!&=\!\mathcal{S}_+[\tilde{\cal I}_{2n,-}]-\mathcal{S}[\tilde{\cal I}_{2n,0}]+\mathcal{S}_+[\tilde{\cal I}_{2n,+}],\quad {\rm Im} \lambda>0,%\label{resum iml p}
        \\
        {\cal I}_{2n}(\lambda)\!&=\!\mathcal{S}_-[\tilde{\cal I}_{2n,-}]+\mathcal{S}[\tilde{\cal I}_{2n,0}]+\mathcal{S}_-[\tilde{\cal I}_{2n,+}],\quad {\rm Im}\lambda<0, %\label{resum iml m},
    \end{aligned}
\end{equation}
which will provide complete information of ${\cal I}_{2n}$.
In the ${\rm Im}(\lambda)\to 0$ limit, the above two combinations will converge to a same expression
\begin{equation}
\begin{aligned}
      {\cal I}_{2n}(\lambda)     =2^{n-\frac{1}{2}} \lambda ^{-\frac{n}{2}-\frac{3}{4}} \left(\sqrt{\lambda } \Gamma
   \left(\frac{n}{2}+\frac{1}{4}\right) \,
   _1F_1\left(\frac{n}{2}+\frac{1}{4};\frac{1}{2};\frac{\sigma ^2}{4 \lambda
   }\right)-\sigma  \Gamma \left(\frac{n}{2}+\frac{3}{4}\right) \,
   _1F_1\left(\frac{n}{2}+\frac{3}{4};\frac{3}{2};\frac{\sigma ^2}{4 \lambda
   }\right)\right),
\end{aligned}
\end{equation}
for non-negative integers $n$.
It is clear that the non-perturbative contribution is related to the perturbative one in this prototype of QFT.

\subsection{Connected correlation function}
We now consider the connected correlation function. By definition, one can express the connected correlation functions by combining the nonconnected ones
\begin{equation}
\begin{aligned}
G_2\!&=\!\gamma_2,\\
G_4\!&=\!\gamma _4-3 \gamma _2^2,\\
G_6\!&=\!30 \gamma _2^3-15 \gamma _4 \gamma _2+\gamma _6,\\
\!&\cdots\!.
\end{aligned}
\end{equation}
Although the analytic form of the right side is known, the analytic form of the connected correlation function is hard to find.

Introducing $V[J]\equiv \frac{\partial}{\partial J}W[J]$, the differential equation (\ref{odeW}) can be rewritten as
\begin{equation}
\lambda \frac{\partial^2}{\partial J^2}V[J]+3\lambda V[J]\frac{\partial}{\partial J}V[J]+\lambda V[J]^3+\sigma V[J]=J
\label{odeV}
\end{equation}
with boundary conditions $V[0]=0$, $V'[0]=G_2$. Given the analytic form of the two-point function $G_2=\gamma_2$, one can obtain the form of any $G_{2n}$ in the power series of
\footnote{This can be done by using the command ``AsymptoticDSolveValue'' in Mathematica.
} 
\begin{equation}
V[J]=\sum_{n=1}^{\infty}\frac{G_{2n}}{(2n-1)!}J^{2n-1},
\label{V expand}
\end{equation}
which is also consistent with the DS equation of the connected functions. 

We then consider the large $n$ asymptotic behavior of $G_{2n}$. This can be done by noting that 
\begin{equation}
  \frac{G_{2n+2}/(2n+1)!}{G_{2n}/(2n-1)!}
   \label{r2}
\end{equation}
approaches a constant $-r^2$ as $n\to \infty$.
We thus can expect $G_{2n}\sim C (-r^2)^n (2n-1)!\quad (n\to \infty)$,
where $C$ is a constant. Comparing with (\ref{V expand}), we get that as $n\to\infty$, $C=-2$. Eventually, we get the asymptotic behavior of the connected correlation function:
\begin{equation}
G_{2n}\sim 2 r^{2n}(-1)^{n+1} (2n-1)!,\quad n\to \infty,
\label{G asym}
\end{equation}
where $r$ is determined by $\sigma$ and $\lambda$. Note that (\ref{G asym}) has the same form as the one in \cite{Bender:2022eze} as $\sigma=0$.
\begin{figure}[H]
    \centering
    \includegraphics[width=0.5\linewidth]{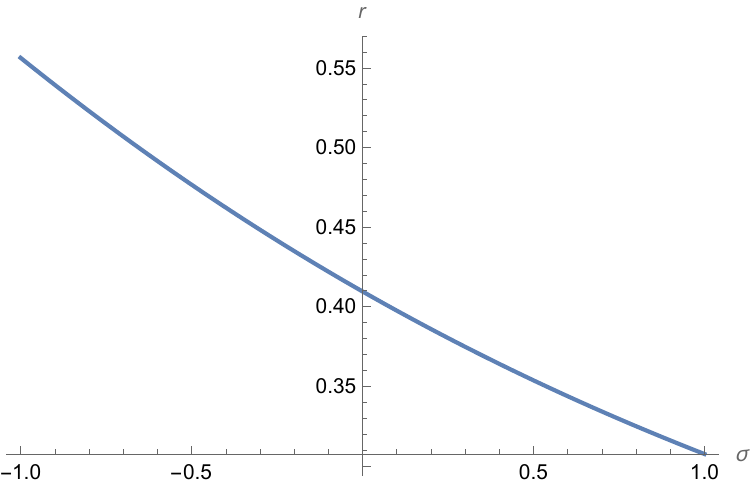}
    \caption{The $\sigma$ dependence of $r$ for $\lambda=1$.}
    \label{r of sigma}
\end{figure}

We then show another way to determine the value of $r$. From the asymptotic behavior, one finds $1/r$ is actually the convergence radius of $V[J]$.  Since $V[J]=\frac{Z'[J]}{Z[J]}$, one can find that $V[J]$ will be divergent as $Z[J]=0$. We first rescale $J= ix$, and the differential equation (\ref{Z ode}) now reads
\begin{equation}
-\sigma \frac{\partial}{\partial x}Y[x]+\lambda \frac{\partial^3}{\partial x^3}Y[x]=x Y[x],
\end{equation}
where $Y[x]=Z[ix]=Z[J]$. With boundary conditions $Y[0]=1$, $Y'[0]=0$ and $Y''[0]=-G_2$, $Y[x]$ can be solved numerically. From the numeric solution, one can find the first zero of $Y[x]$, which is nothing but the convergence radius $1/r$. In Fig.\ref{r of sigma}, we show how $r$ depends on $\sigma$, which is continuous as $\sigma=0$.

\section{Truncation of the DS Equations}
\label{sec:truncation}

In this section, we consider the truncation of the DS equation based on the exact results obtained in the previous section. To gain insights into more general QFT, we will start with the asymptotic expansion around the perturbative saddle point, which is analogous to the one obtained from the Feynmann diagram in general QFT. Since the asymptotic expansion is divergent, we will consider its Borel resummation instead. We will consider the truncation for nonconnected and connected correlation functions separately.  

\subsection{Truncation for nonconnected correlation function}

\begin{figure}[H]
\begin{center}
 \begin{subfigure}{0.32\textwidth}
\includegraphics[width=\linewidth]{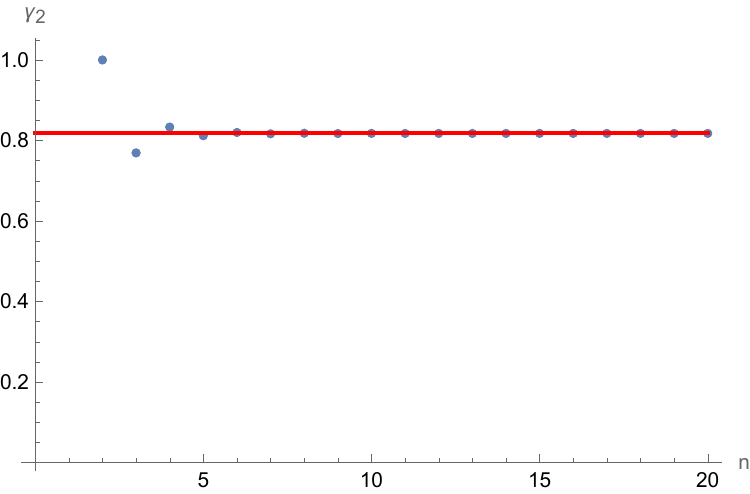}
    \caption{}
    \end{subfigure}
\begin{subfigure}{0.32\textwidth}
\includegraphics[width=\linewidth]{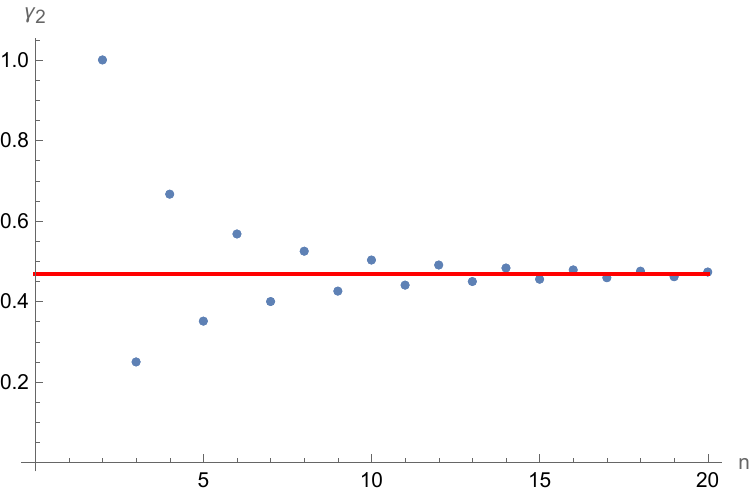}
    \caption{}
    \end{subfigure}
\begin{subfigure}{0.32\textwidth}
\includegraphics[width=\linewidth]{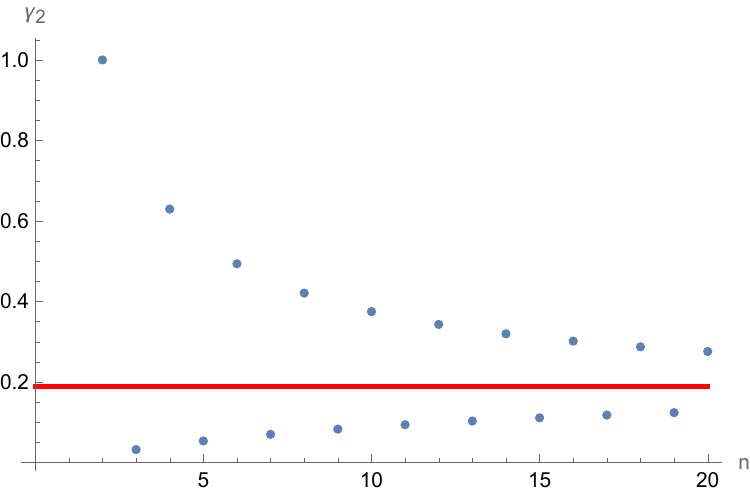}
    \caption{}
    \end{subfigure}\\
    \begin{subfigure}{0.32\textwidth}
 \includegraphics[width=\linewidth]{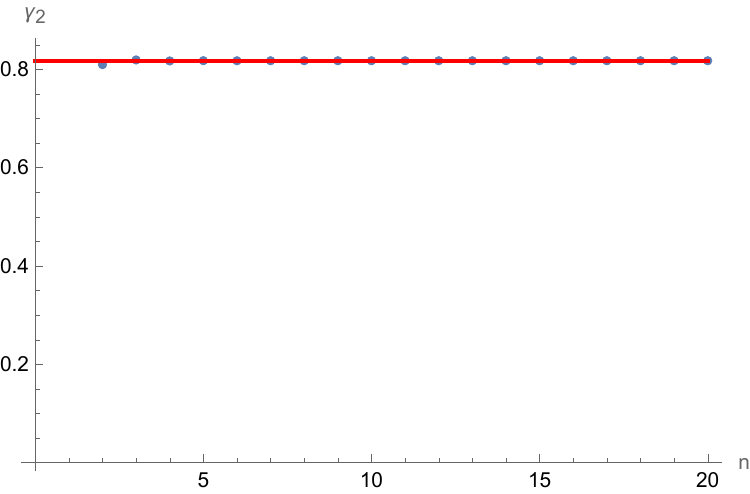}
    \caption{}     
    \end{subfigure}
    \begin{subfigure}{0.32\textwidth}
 \includegraphics[width=\linewidth]{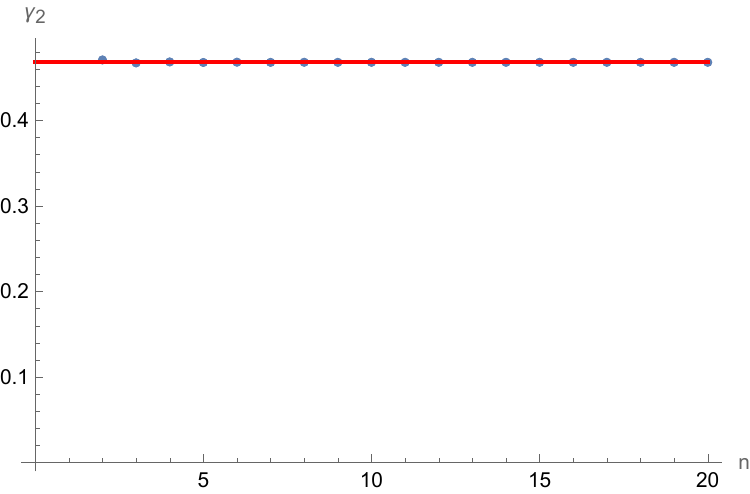}
    \caption{}     
    \end{subfigure} 
    \begin{subfigure}{0.32\textwidth}
 \includegraphics[width=\linewidth]{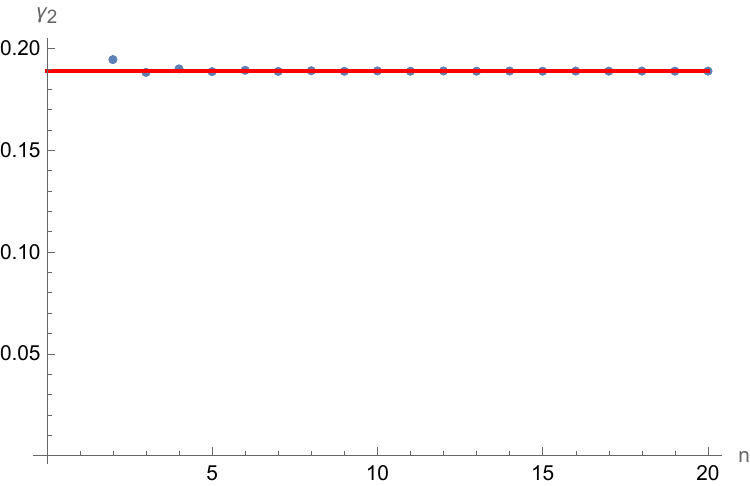}
    \caption{}     
    \end{subfigure}
    \end{center}
    \caption{In this figure we let $\sigma=1$. The red lines in the figure represent the exact values of $\gamma_2$ calculated using the integral form (\ref{gamma 2n}), while the blue dots correspond to the solutions obtained by truncating the DS equations at $\gamma_{2n}$. The upper panels show the traditional truncation with $Q_{2n}(\gamma_2)=0$, and the lower panels show the truncation using \eqref{sigmapasym}. 
    %by letting $Q_{2n}(\gamma_2)$ be the large $n$ asymptotic behavior of ${\cal I}_{2n}(\lambda)$ up to subleading order. 
    From the left panels to the right panels, $\lambda=0.1, 1,10$ respectively.}
    \label{trunc sigma>0}
\end{figure}

\paragraph{Truncation: $\sigma>0$}
The traditional way to truncate the DS equation is to set $\gamma_{2n}=0$ for a sufficiently large $n$. To test this approximation, we express $\gamma_{2n}$ as a function of $\gamma_2$, i.e. $\gamma_{2n}=Q_{2n}(\gamma_2)$, using the DS equation \eqref{eq:DS-non-dis}. In the upper panels of Fig.\ref{trunc sigma>0}, we show the real zeros (blue dots) of  $Q_{2n}(\gamma_2)$ for $n=2,\cdots,20$, and compare them with the exact result of $\gamma_2$ (red lines in Fig.\ref{trunc sigma>0}). As shown in Fig.\ref{trunc sigma>0}, the traditional truncation is inappropriate for the strong coupling. Even at weak coupling, the accuracy of the approximation is low.

To improve the precision, we propose to truncate the DS equation by using the large $n$ asymptotic behavior of ${\cal I}_{2n}(\lambda)$, which can be obtained by using the large $a$ expansion of $U(a,c,z)$\cite{doi:10.1142/9195}
\begin{equation}
    U(a,c,z)\sim 2\left(\frac{z}{a}\right)^{\frac{1}{2}(1-c)}\frac{e^{\frac12z}}{\Gamma(a)}\sum_{k=0}^{\infty}c_k(z)\left(\frac{z}{a}\right)^{\frac{1}{2}k}K_{c-k-1}(2\sqrt{az}),
\end{equation}
where coefficients $c_k(z)$ are given by
\begin{equation}
     c_0(z)=1,\quad  c_1(z)=\frac{1}{2}(6c-z),\quad 
    c_2(z)=\frac{1}{288}(-12c+36c^2-12zc+z^2),\quad\cdots\,.
\end{equation}
%\eqref{SAlambda sigma>0}:
Together with the expansion of Gamma function, one finds the expansion of $ {\cal I}_{2n}$ up to the subleading order
\begin{equation}
    \begin{aligned}
        {\cal I}_{2n}\sim&\frac{1}{27 \lambda  (2 n+1)^2}\sqrt{\pi } 2^{n-\frac{11}{2}} n^{n-1} (3 n+1) (24 n-1) (\lambda +2 \lambda 
   n)^{\frac{1}{4} (-2 n-3)} \\
   &\times\Big(12 \lambda  \big(2 n (4 \sqrt{\lambda +2
   \lambda  n}+\sigma )+6 \sqrt{\lambda +2 \lambda  n}+\sigma \big)\\
   &-\sigma ^2
   (2 \sqrt{\lambda +2 \lambda  n}+2 n \sigma +\sigma )\Big) e^
   {\frac{-4 \sigma  \sqrt{\lambda +2 \lambda  n}+\lambda  (2-4 n)+\sigma ^2}{8
   \lambda }}.
    \end{aligned}
    \label{sigmapasym}
\end{equation}
In the lower panels of Fig.\ref{trunc sigma>0}, we truncate the DS equation by approximating $Q_{2n}(\gamma_2)$ with the large $n$ expansion of ${\cal I}_{2n}(\lambda)$ up to the subleading order. As presented in Fig.\ref{trunc sigma>0}, our truncation yields a significant improvement. More remarkably, our truncation is also suitable for the strong coupling regime.

% \begin{figure}[H]
% \begin{center}
%  \begin{subfigure}{0.38\textwidth}
% \includegraphics[width=\linewidth]{gm2n-n20-S1-L0-1-0.pdf}
%     \caption{}
%     \end{subfigure}
%     \begin{subfigure}{0.38\textwidth}
%  \includegraphics[width=\linewidth]{gm2n-n20-S1-L0-1-large-n.pdf}
%     \caption{}     
%     \end{subfigure}\\
% \begin{subfigure}{0.38\textwidth}
% \includegraphics[width=\linewidth]{n20-sigma1-lambda1-0-gamma2n.pdf}
%     \caption{}
%     \end{subfigure}
%     \begin{subfigure}{0.38\textwidth}
%  \includegraphics[width=\linewidth]{n20-sigma1-lambda1-largen1-gamma2n.pdf}
%     \caption{}     
%     \end{subfigure}\\  
%     \begin{subfigure}{0.38\textwidth}
% \includegraphics[width=\linewidth]{gm2n-n20-S1-L10-0.pdf}
%     \caption{}
%     \end{subfigure}
%     \begin{subfigure}{0.38\textwidth}
%  \includegraphics[width=\linewidth]{gm2n-n20-S1-L10-large-n.pdf}
%     \caption{}     
%     \end{subfigure}
%     \end{center}
%     \caption{In this figure we let $\sigma=1$. The red lines in the figure represent the exact values of $\gamma_2$ calculated using the integral form (\ref{gamma 2n}), while the blue dots correspond to the solutions obtained by truncating the DS equations at $\gamma_{2n}$. The left panels show the traditional truncation with $Q_{2n}(\gamma_2)=0$, and the right panels show the truncation by letting $Q_{2n}(\gamma_2)$ be the large $n$ asymptotic behavior of ${\cal I}_{2n}(\lambda)$ up to subleading order. From the upper panels to the lower panels, $\lambda=0.1, 1,10$ respectively.}
%     \label{trunc sigma>0}
% \end{figure}

\begin{figure}[H]
   \begin{center}
 \begin{subfigure}{0.48\textwidth}
\includegraphics[width=\linewidth]{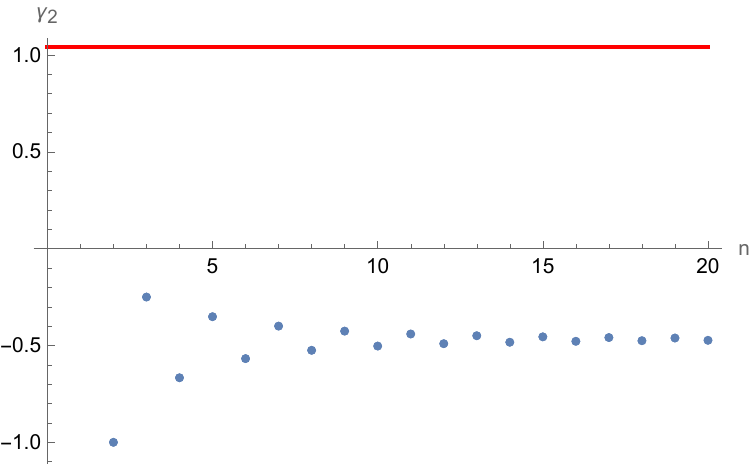}
    \caption{}
    \end{subfigure}
\begin{subfigure}{0.48\textwidth}
\includegraphics[width=\linewidth]{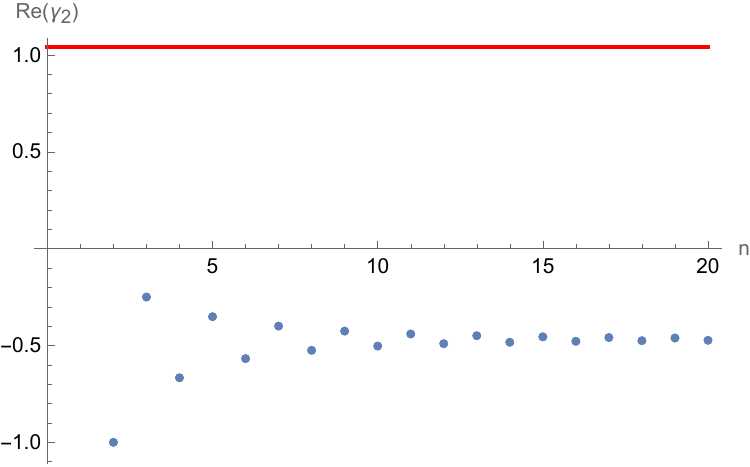}
    \caption{}
    \end{subfigure}
    \begin{subfigure}{0.48\textwidth}
\includegraphics[width=\linewidth]{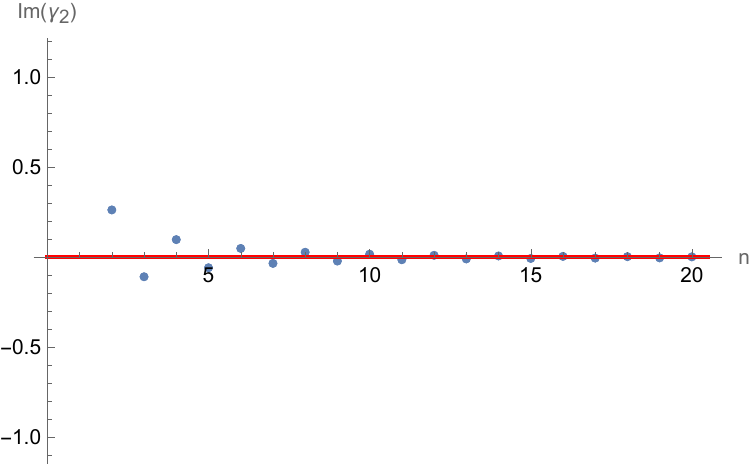}
    \caption{}
    \end{subfigure}
\begin{subfigure}{0.48\textwidth}
\includegraphics[width=\linewidth]{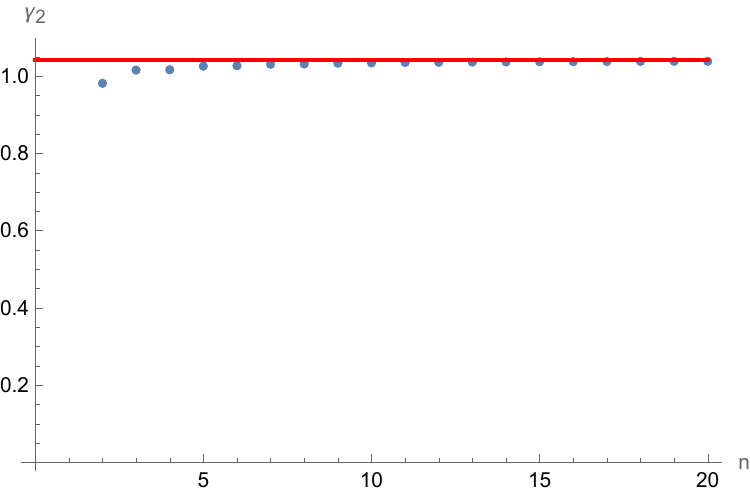}
    \caption{}
    \end{subfigure}
    \end{center}
    \caption{In this figure $\sigma=-1$, $\lambda=1$. $(a)$ shows the traditional truncation with $Q_{2n}(\gamma_2)=0$. $(b)$ and $(c)$ show the truncation with $Q_{2n}(\gamma_2)={\cal S}[{\cal I}_{2n,0}]$ (\ref{x0 resum sigma m}). 
    The blue dots in $(b)$ represents the real parts of the solutions. $(c)$ shows the imaginary parts of the solutions, which approach $0$ as $n$ increases. $(d)$ shows the truncation by replacing $Q_{2n}(\gamma_2)$ by the large $n$ asymptotic behavior of ${\cal I}_{2n}$ \eqref{sigmanasym}.}
     \label{trunc sigma<0}
\end{figure}

\paragraph{Truncation: $\sigma<0$}
In the case of $\sigma<0$, the traditional truncation is much worse even in the weak coupling regime, see $(a)$ in Fig.\ref{trunc sigma<0}. One natural guess to improve the truncation is to use the Borel resummation ${\cal S}[{\cal I}_{2n,0}]$ \eqref{x0 resum sigma m}, where only the perturbative saddle point is considered. However, as shown in $(b)$ and $(c)$ in Fig.\ref{trunc sigma<0}, this truncation method does not enhance the accuracy. Those two kinds of truncation appear quite similar, because ${\cal S}[{\cal I}_{2n,0}]$ is close to zero at large $n$ in this case.

To address this issue, we will truncate the DS equation using the large $n$ asymptotic behavior of ${\cal I}_{2n}=\mathcal{S}_\pm[{\cal I}_{2n,-}]\mp\mathcal{S}[{\cal I}_{2n,0}]+\mathcal{S}_\pm[{\cal I}_{2n,+}]$ \eqref{resum iml pm}. This can be done using the large $a$ expansion of $\,_1F_1(a,c,z)$ expansion \cite{doi:10.1142/9195}:
\begin{equation}
    \frac{1}{\Gamma(c)} \,_1F_1(a;c;z)\sim a^{c-1}\frac{\Gamma(1+a-c)e^{\frac12z}}{\Gamma(a)}\sum_{k=0}^{\infty}(-1)^k\frac{c_k(z)}{a^k}E_{c-k-1}(-az),
\end{equation}
where $E_\nu(z)$ can be expressed using the Bessel function of the first kind $E_\nu(z)=z^{-\frac{1}{2}\nu}J_\nu(2\sqrt{z})$. Applying these to ${\cal I}_{2n}$, one finds
\begin{equation}
\begin{aligned}
    & {\cal I}_{2n}\sim\frac{1}{9
   (\lambda  n)^{7/4}}\sqrt{\pi } 2^{n-\frac{33}{4}} n \lambda ^{-n/2} e^{\frac{1}{8} \left(\frac{\sigma ^2}{\lambda }+4 n \left(\log \left(\frac{n}{2}\right)-1\right)\right)}\Bigg\{\frac{24 (4 n+1) \sqrt{\lambda  n (4 n+2)}}{(2 n+1)^{5/2}}\\
   &\times\Bigg[\left(24 \lambda  (4 n+3)-2 \sigma ^2\right) \cosh \left(\frac{1}{2} \sigma  \sqrt{\frac{2 n+1}{\lambda }}\right)+\sigma  \left(\sigma ^2 \sqrt{\frac{2 n+1}{\lambda }}-12 \sqrt{\lambda +2 \lambda  n}\right) \sinh \left(\frac{1}{2} \sigma  \sqrt{\frac{2 n+1}{\lambda }}\right)\Bigg]\\
   &-\frac{(4 n+9) (48 n-1) \left(\left(\sigma ^3-36 \lambda  \sigma \right) \cosh \left(\frac{1}{2} \sigma  \sqrt{\frac{2 n+3}{\lambda }}\right)+48 \lambda 
   \sqrt{\lambda  (2 n+3)} \sinh \left(\frac{1}{2} \sigma  \sqrt{\frac{2 n+3}{\lambda }}\right)\right)}{(2 n+3)^2}\Bigg\},
\end{aligned}
\label{sigmanasym}
\end{equation}
which is up to subleading order. In $(d)$ of Fig.\ref{trunc sigma<0}, we let $Q_{2n}(\gamma_2)$ be the large $n$ expansion of ${\cal I}_{2n}$ upto subleading order\eqref{sigmanasym}. This approach shows a significant improvement in this case. 

This suggests that, although the asymptotic series around the perturbative saddle point is Borel summable, it does not capture the full information. We have to include the contributions of the non-perturbative saddle points in the truncation procedure, which should shed light on the truncation procedure in general QFT.

\subsection{Truncation for Connected Correlation Functions}
In this subsection, we truncate the DS equations \eqref{eq:DS-conn} for the connected correlation functions by using the large $n$ asymptotic behavior \eqref{G asym}.

\begin{figure}[H]
\begin{subfigure}{0.5\textwidth}
\includegraphics[width=\linewidth]{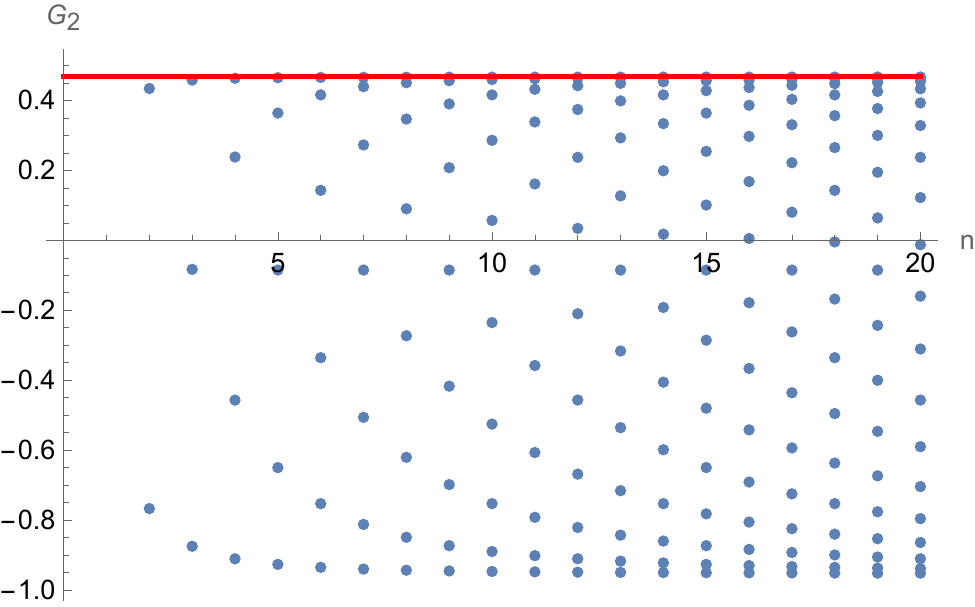}
    \caption{}
    \end{subfigure}
    \begin{subfigure}{0.5\textwidth}
 \includegraphics[width=\linewidth]{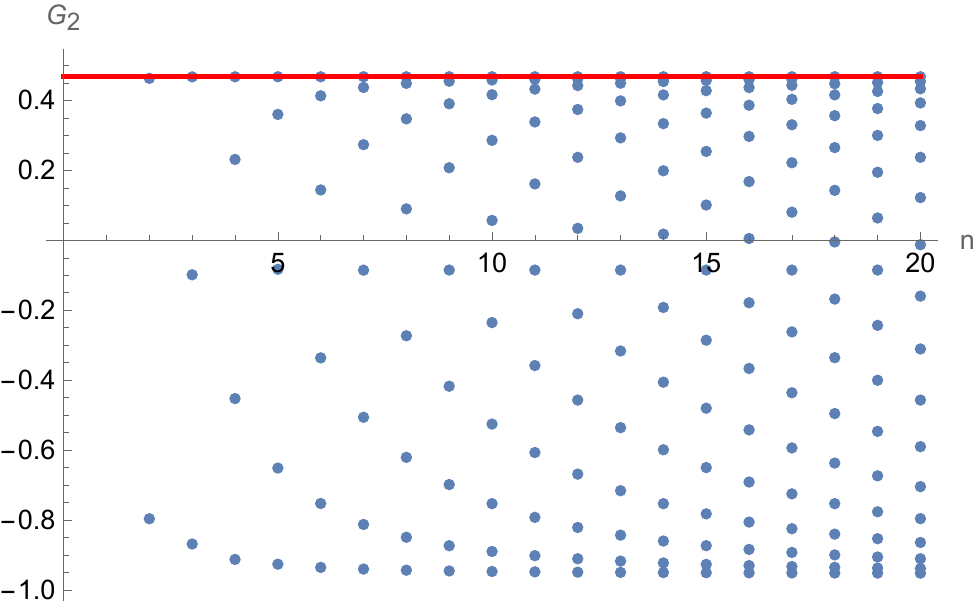}
    \caption{}     
    \end{subfigure}  
    \label{fig:enter-label}
    \caption{$\sigma=1$, $\lambda=1$, $r=0.30737887\cdots$. The left figure shows the truncation with $P(G_2)=0$, and the right figure shows the truncation by using the asymptotic behavior of the connected correlation function (\ref{G asym}). A similar result is obtained for different values of $\lambda$.}
    \label{G trunc sigma>0}
\end{figure}

\paragraph{Truncation: $\sigma>0$}
We begin by considering the case where $\sigma>0$. We express $G_{2n}$ in terms of $G_2$, and denote it by $P_{2n}(G_2)=G_{2n}$. In our truncation, we let $P_{2n}(G_2)\sim 2r^{2n}(-1)^{n+1}(2n-1)!$, whose real solutions $G_2$ (blue dots) is shown in $(b)$ of Fig.\ref{G trunc sigma>0}\footnote{See \cite{Bender:2022eze} for the case $\sigma=0$.}. $(a)$ of Fig.\ref{G trunc sigma>0} shows the real zeros (blue dots) of $P_{2n}(G_2)$, which is the result of the traditional truncation method. 
%For $\sigma=1$ and $\lambda=1$, $r=0.30737887\dots.$

\begin{figure}[H]
\begin{subfigure}{0.5\textwidth}
\includegraphics[width=\linewidth]{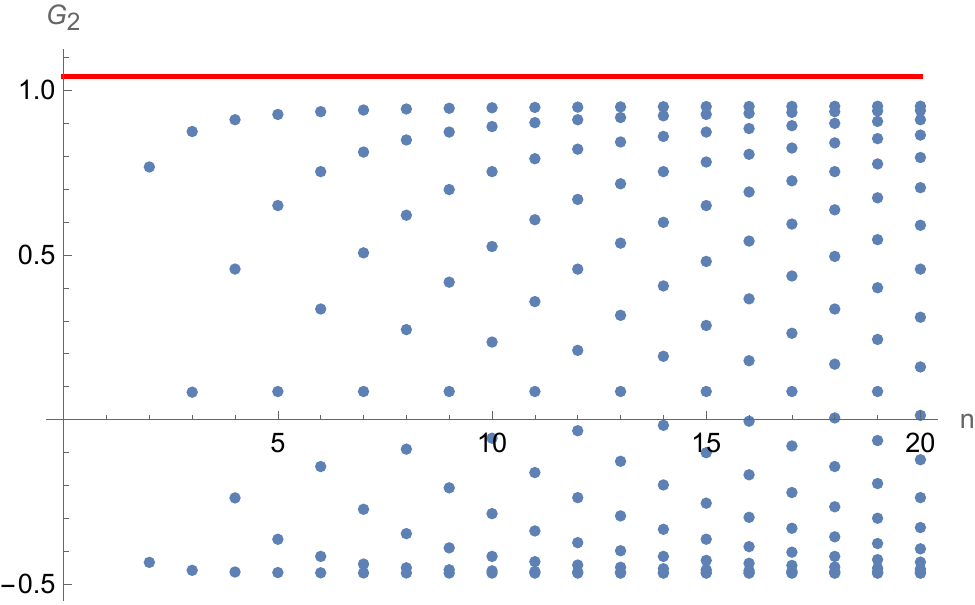}
    \caption{}
    \end{subfigure}
    \begin{subfigure}{0.5\textwidth}
 \includegraphics[width=\linewidth]{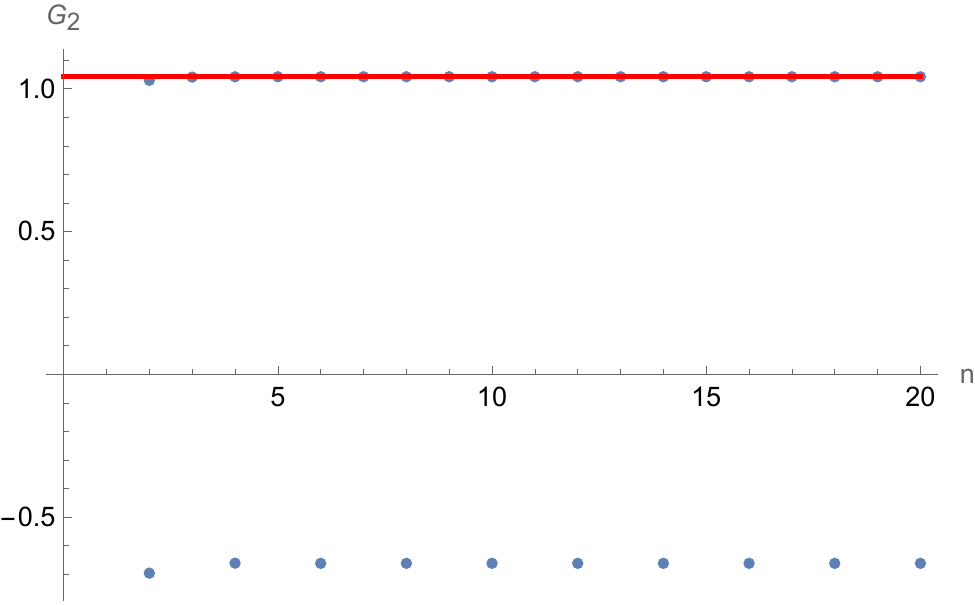}
    \caption{}     
    \end{subfigure}  
    \caption{$\sigma=-1$, $\lambda=1$, $r=0.55631867\cdots$. The left figure shows the truncation with $P(G_2)=0$, and the right figure shows the truncation by using the asymptotic behavior of the connected correlation function (\ref{G asym}).}
    \label{G trunc sigma<0}
\end{figure}

\paragraph{Truncation: $\sigma<0$}
When $\sigma<0$. The story is analogous. In $(a)$ of Fig.\ref{G trunc sigma<0}, the real zeros (blue dots) of $P_{2n}(G_2)$ are shown. The real solutions  (blue dots) of $P_{2n}(G_2)\sim 2r^{2n}(-1)^{n+1}(2n-1)!$ are shown in $(b)$ of Fig.\ref{G trunc sigma<0}.

Obviously, our truncation method provides a significantly higher accuracy compared to the traditional approach. This is more remarkable in the case $\sigma<0$.

\section{Conclusion and discussion}
\label{sec:conclusion}

In this work, we apply the Lefschetz
thimble decomposition and the saddle point expansion to the correlation functions in zero-dimensional QFT. From these asymptotic formal series, we have reconstructed the exact correlation function using the Borel resummation. To gain insight into the general QFT, we consider the truncation procedure of the DS equations starting with the asymptotic expansion around the perturbative saddle point. Unlike the traditional truncation method, which is suitable only for the weak coupling regime and offers limited accuracy, our proposed truncation method addresses these shortcomings and demonstrates a significant improvement, even in the case of the strong coupling regime. When $\sigma>0$, the perturbative series is Borel summable, and the truncation based on the Borel resummation of the perturbative saddle point is sufficient. However, when $\sigma<0$, the situation becomes more complicated. Although the asymptotic series around the perturbative saddle point is Borel summable, it does not capture the full information\footnote{This implies that the truncation based on the Feynmann diagram perturbation may lose information in some cases, even the asymptotic expansion is Borel summable.}. In this case, it is necessary to include the non-perturbative saddle point. In the truncation procedure, these non-perturbative contributions must be taken into account. In essence, we should take advantage of the asymptotic behavior of the full trans-series, incorporating both perturbative and non-perturbative contributions. We believe our findings will provide valuable insights for the study of more general QFTs.

It would be interesting to consider the DS equations of higher dimensional QFT or QM, where the Borel resummation is also necessary. However, due to the complexities associated with renormalization, this is a hard task, but an interesting direction to explore. Some attempts on QM can be found in \cite{Li:2023nip,Li:2023ewe,Li:2024rod}. It is also interesting to explore a more complicated zero-dimensional model, such as the PT-symmetric QFT \cite{Bender:1999ek}, where modification to the integral contour should be considered. See also \cite{Ai:2022csx} for related topics.

\subsection*{Acknowledgements}
We would like to thank Yong Cai, Jie Gu, Wenliang Li, Yong Li, Zhibin Li, Hao Ouyang, Zhaojie Xu and Hao Zou for useful discussions. H.S. is supported by the National Natural Science Foundation of China No.12405087 and the Startup Funding of Zhengzhou University (Grant No.121-35220049). H.S. would like to thank ``Resurgence Theory in Mathematical Physics,'' ``Tianfu Fields and Strings 2024'' and ``Gravity, Field Theories, and Their Relations 2024 -- Integrability: From Mathematics to Experiments'' for their support and warm hospitality.

\bibliographystyle{ytphys.bst}
\bibliography{BIBLIOGRAPHY}

\end{document}